\documentstyle[11pt,newpasp,twoside,epsf]{article}
\def\edcomment#1{\iffalse\marginpar{\raggedright\sl#1\/}\else\relax\fi}
\reversemarginpar

\begin{document}
\title{Host galaxies of AGN}
 \author{C. Boisson, M. Joly}
\affil{Observatoire de Paris-Meudon, F-92195 Meudon Cedex}

\begin{abstract}
  
  The relationship of an AGN to its host galaxy is a crucial
  question in the study of galaxy evolution. We perform stellar
  population synthesis in the central regions of galaxies of different
  activity levels. A large number of stellar features are measured
  both in the optical and near-infrared.
  
  We find the nuclear stellar population to be related to the level of
  activity. These differences are no more conspicuous further away in
  the bulge of the galaxy.

\end{abstract}

\section{Introduction}

 An unsolved issue in understanding the Seyfert phenomenon is the
 relation between the Seyfert type of activity and its host galaxy. A
 number of studies have found a connection between nuclear activity
 and strong star formation in the host galaxy.  Studying host galaxy
 of Active Galactic Nuclei (AGN) can help answering questions relevant
 to the understanding of AGN but also to galaxy formation and
 evolution such as : how does the AGN affect the host galaxy? Is there
 a direct link between the activity level of the AGN and the starburst
 activity? Is the starburst activity nuclear or circum-nuclear?  What
 about the age and metallicity of the host galaxy stellar population.

In order to answer to some of these questions, we perform stellar
population synthesis of the integrated starlight emitted in the
central 10 arcsec of nearby Active Galaxies in the optical and near-IR
range. The good spatial resolution now achievable allows to detect
gradients of population and starburst activity in the central region
of the galaxies.

\section{Observations}
Long slit spectroscopy of a sample of AGN, including 5 Seyfert~1, 5
Seyfert~2, 3 LINERs and 2 Starburst galaxies was obtained in the
visible; 2 Seyfert~1 and 3 Seyfert~2 were also observed in the H band.

In the optical, observations were obtained with the Herzberg at CFHT
(Serote Roos et al., 1998) and with EMMI at NTT, in the range
5000-9000\AA\ with a resolution of FWHM=10\AA. Spectra are extracted
from the central 3-4~arcsec, as well as from surrounding regions
located at 5 to 7~arcsec from the center, i.e. within the bulge at the
distance of the galaxies. Up to 47 stellar features were measured.
These spectra are compared to a library of star spectra (see
Section 2).

In the near-infrared domain there is a much better contrast of galaxy
light to nuclear light. This domain allows to view the host galaxy
with less contamination from the nucleus than is possible in the
visible. Moreover cool stellar populations are best studied as their
spectral energy distribution peaks near 1$\mu$. Also it allows to
probe obscured objects since extinction effects are much less
important at these wavelengths. The H band was favoured as it is
remarkably rich in strong metallic features and exhibits powerful
luminosity indicators(Dallier et al.,1996). ISAAC at VLT, in Medium
Resolution mode, provides a resolution of FWHM=5\AA. In the range
1.56-1.64$\mu$m, 30 metallic stellar features are identified, among
which some line ratios are strongly dependent on the luminosity class
of the stars, while quasi no emission lines from the AGN are
present (Fig.~1). A flux calibrated stellar library at a resolution
R=2000 is available (Dallier et al., 1996) and is being extended for a
better coverage of the H-R diagram.

\begin{figure}
\vspace{7cm}
\includegraphics{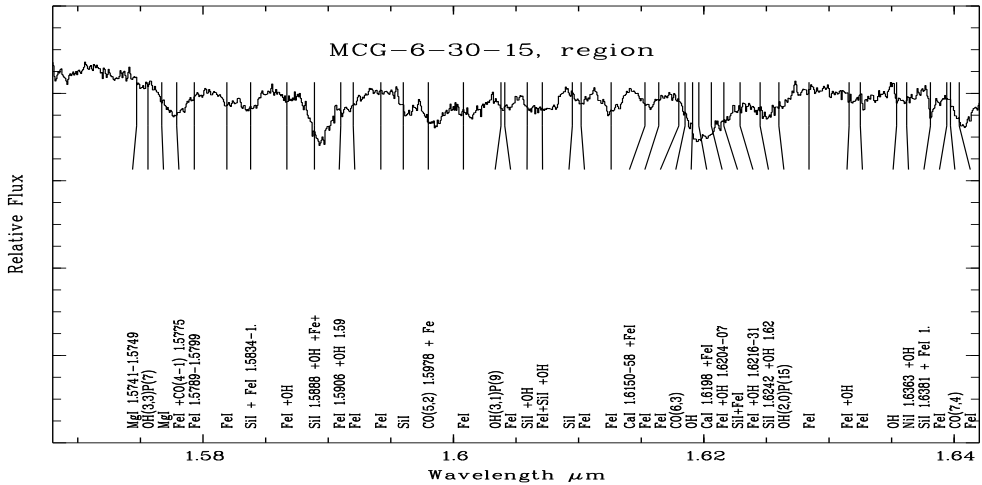}
\caption{H band spectrum at a resolution R=3200 of a circumnuclear region 
of the Seyfert~1 MGC-6-30-15 at 1.5'' from the nucleus. 
Numerous metallic stellar features can be identified.}
\end{figure}

\section{Synthesis.}

There are two ways of computing stellar population synthesis. The {\it
  direct} approach presupposes perfect knowledge of the stellar
evolution in order to create appropriate model to a composite
population with specified age, metallicity and star formation rate.
The {\it inverse} approach where populations are synthesized in terms
of {\it observables}. In this approach, the one we favour, stellar
evolution is used only to introduce constraints as to lower the number
of possible solutions and give more weight to the adopted solution.

To compute stellar population synthesis, we compare the equivalent
width (EW)of numerous stellar features observed in the galaxy
to the EW of the same lines measured in a library of star spectra.
Recall that EW are independent of reddening.

We get a system of non-linear equations solved using a method
developped by Pelat (1997 and 1998) and Moultaka and Pelat (2000)
which determines the best solution. The advantage of this method is
that no hypothesis on the IMF, or on the history of star formation is
made, and no evolutionary track models are preferred. It gives the
contribution of each stellar type to the total radiation at a
reference wavelength and the standard deviation to these
contributions.

Synthetic spectra are computed using the stellar library. 
Intrinsic reddening is deduced from the comparison 
of the overall shape of the observed and synthetic spectra. 

An example of the results obtained in the optical range is shown
Figure~2 where the synthetic spectrum of each region extracted from
one of the LINER galaxies are superposed to the observed spectrum
after correction of the internal reddening.

\begin{figure}[t]
\vspace{7cm}
\includegraphics{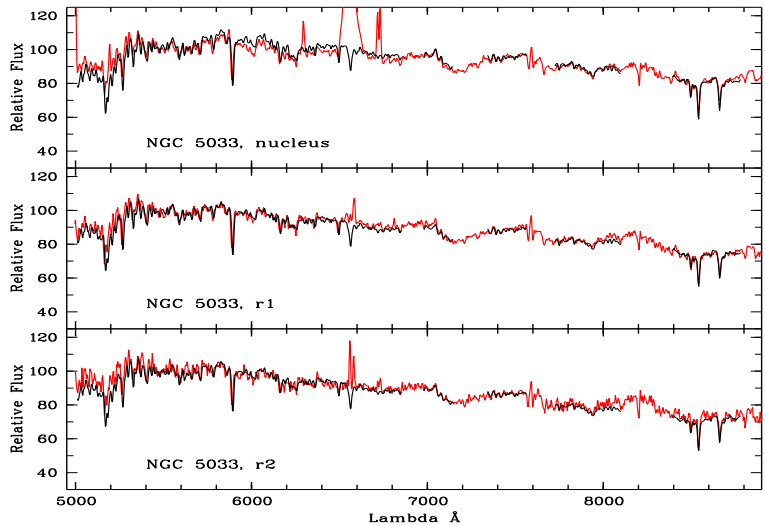}
\caption{Synthetic spectra (black line) superimposed on the observed one 
(grey line)}
\end{figure}

Such a study in the optical lead to two important results i) the
stellar populations within a class of AGN activity, whatever the
morphological type of the host galaxy, is very homogeneous and ii)
the populations are different in the very central regions of
Seyfert~2 and LINERs (Boisson et al., 2000 \& 2001). LINERs have a
very old metal rich population while in Seyfert~2 a contribution of
weak or old burst of star formation is observed in addition to the old
high metallicity component (in comparison, Starbursts have a stronger
and younger stellar population).

In the infrared, the work is still in progress. But some hints of the
power of combining optical and infrared bands can be given.  Figure~3
displays the spectrum of a circumnuclear region, located at 360pc from
the center, of the Seyfert~1 galaxy MCG-6-30-15. The spectral features
are similar to those in the nucleus (within a radius R=180pc) although
quite different in strength, indicating that the stellar population
may be the same in both regions but that in the nucleus stellar
features are diluted by a continuous emission. This diluting continuum
may be due to dust emission, in which case if one assumes no gradient
of population, the comparison of the overall shape of the spectra 
indicate a dust temperature of about 1000K.

\begin{figure}
\vspace{7cm}
\includegraphics{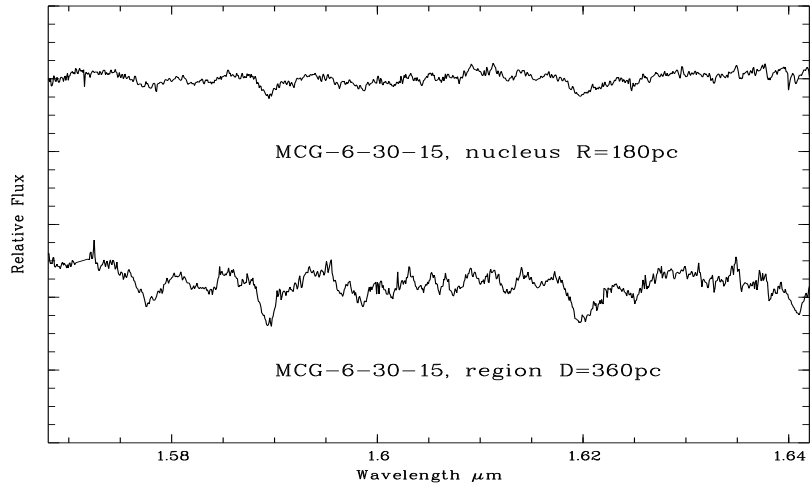}
\caption{H band spectra of MCG-6-30-15}
\end{figure}

\begin{figure}
\vspace{7cm}
\includegraphics{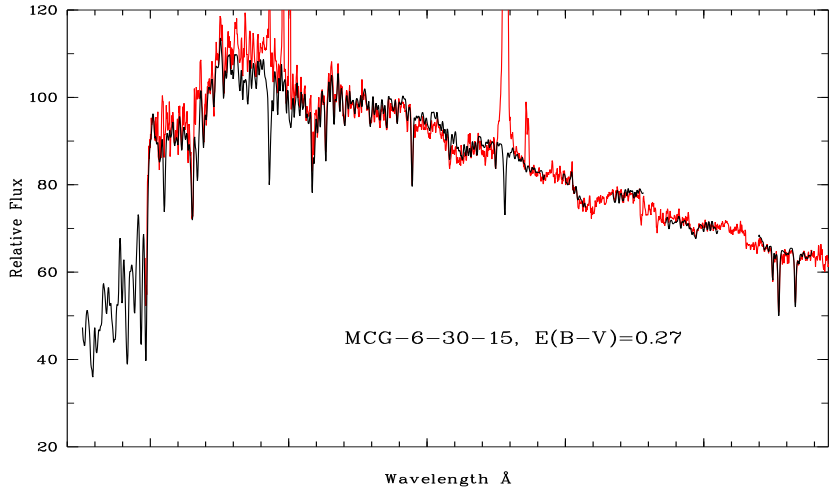}
\caption{Synthetic spectrum of MCG-06-5-30 (black line) superimposed 
  on the observed optical one (grey line).}
\end{figure}

Figure~4 shows the optical spectrum of the same region with the
synthetic spectrum superimposed. The stellar population synthesis
indicates the presence of 67\% dwarf and 23\% giant stars. At
1.6$\mu$m this translates into 12\% dwarf and 88\% giant stars with
60\% M giants and 20\% K giants. In Figure~5 we show this
``translated'' synthetic stellar population diluted by 20\% host dust
on top of the nuclear spectrum of MCG-6-30-15. The comparison is quite
encouraging.

\begin{figure}
\vspace{7cm}
\includegraphics{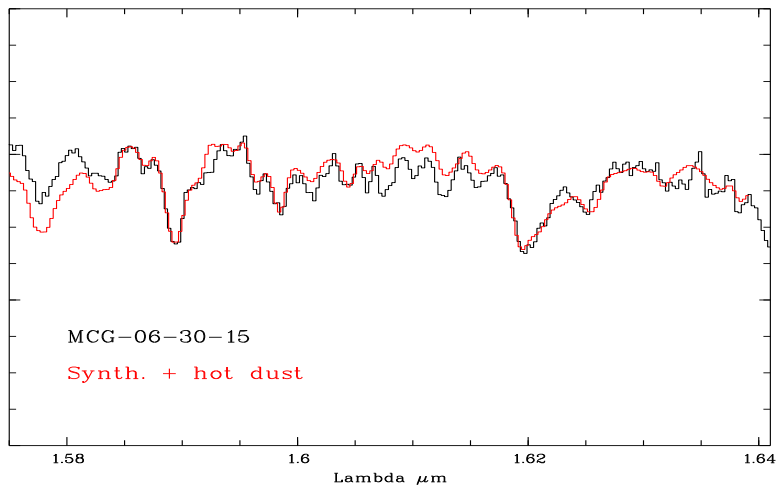}
\caption{``Translated''
  optical synthesis spectrum (grey line) superimposed on the nuclear
  infrared spectrum of MCG-06-5-30 (black line)}
\end{figure}

\section{Conclusions.}

The stellar population in the inner regions of AGN are related to the
level of activity of the AGN. These differences are no more
conspicuous further away in the bulge of the galaxies. A relationship
between the nucleus activity and circumnuclear starburst is advocated.

The near-infrared domain allows to study as well the very nucleus of
Seyfert~1 as evidenced in the case of MCG-6-30-15. Future full
spectral synthesis in the IR spectra will be performed.

\end{document}